\documentclass[12pt]{iopart}
\usepackage[dvips]{graphics, color}
\usepackage{iopams,wrapfig}
\usepackage{xspace}
\usepackage{subfig}
\usepackage{graphicx}
\usepackage{cite}

\begin{document}
\title[The Onset of Jet Quenching in p-p collisions]{The Onset of Jet Quenching Phenomenon}

\author{M. T. AlFiky$^1$, O. T. ElSherif$^1$ and A. M. Hamed$^{1,2,3}$}


\address{$^{1}$ Department of Physics, The American University in Cairo, New Cairo 11835, Egypt}
\address{$^2$ Department of Physics \& Astronomy, Texas A\&M University, College Station, TX 77843, USA}
\address{$^3$ Department of Physics \& Astronomy, University of Mississippi, Oxford, MS 38677, USA}

\eads{\mailto{alfiky@aucegypt.edu}, \mailto{omar343@gmail.com}, \mailto{ahamed@comp.tamu.edu}}


\begin{abstract}
The aim of this study is to set a baseline for the jet quenching measurements of the Quark Gluon Plasma (QGP) formed in the large system size Nucleus-Nucleus (A-A) at top central collisions, via studying simulated small system size, Nucleon-Nucleon (N-N) collisions. The proton-proton (p-p) collisions were simulated using PYTHIA, at center of mass energies $\sqrt{s_{_{NN}}} = 200$ $GeV$ and $\sqrt{s_{_{NN}}} = 13$ $TeV$ corresponding to the available energies at the current collider experiments; the Relativistic Heavy Ion Collider (RHIC), and the Large Hadron Collider (LHC). At both energies, the two-particle azimuthal correlation functions have been considered, and the yield associated with the high transverse momentum ($p_{_{T}}$) particles were extracted at its near-side ($\Delta\phi \approx 0$) and away-side ($\Delta\phi \approx \pi$) at mid pseudo rapidity ($|\eta| \le 2$). The ratio between the near-side yields in the high multiplicity events to these of the low multiplicity events ($I_{_{HL}}^{^{N}}$), as well as, the ratio of the away-side yields ($I_{_{HL}}^{^{A}}$) were calculated at both energies as a function of the hadron fractional energy $z_{_{T}}$ of the high-$p_{_{T}}$ particle.
At both energies, the values of $I_{_{HL}}^{^{N}}$ and $I_{_{HL}}^{^{A}}$ were less than unity, and of trivial dependence on $z_{_{T}}$. The values of $I_{_{HL}}^{^{A}}$ are always less than these of $I_{_{HL}}^{^{N}}$ at the same multiplicity and energy, and both quantities show a pattern of systematic decreases with the multiplicity. Such multiplicity dependence cannot be used neither to exclude the jet quenching nor to prove it in the high multiplicity events in p-p collisions, as the suppressions have been found at both sides, near and away of the high-$p_{_{T}}$ particle.
\end{abstract}

\section{Introduction}
Many interesting results from RHIC and LHC have not only indicated the
QGP formation in the central A-A collisions (events with high multiplicity), but also reflected the intricacy and richness of the Quantum Chromodynamics (QCD). Among the high-$p_{_{T}}$ observables, the suppression of the spectra of the strongly
interacting particles (hadrons) in central A-A collisions compared to the p-p collisions and to the peripheral A-A collisions at LHC and RHIC suggests the strong interaction between the propagated partons and the
formed QGP medium in central A-A collisions \cite{1,2,3,4,5}. These results have been confirmed with the similar spectra of the electromagnetic interacting particles (direct photons) and
the weakly interacting particles (Z$^{0}$ and W$^{\pm}$) produced in central A-A collisions compared to those in p-p
collisions \cite{6,7,8,9}. Furthermore, the two-particle azimuthal correlation measurements have shown the suppression of away-side yield per trigger particle in central A-A collisions compared to the peripheral A-A collisions and p-p collisions \cite{1}. Combining these sets of results stipulates the formation of the QGP in the central A-A collisions, and the viability of the commonly used observables e.g.; the nuclear modifications factor, etc.. ($R_{_{AA}}$, $R_{_{CP}}$, $I_{_{AA}}$, $I_{_{CP}}$). However, the absence of different levels of suppression for light vs. heavy quarks \cite{10,11}, and quark vs. gluon jets \cite{12}, in contrast to the QCD predictions, has necessitated the need for more sensitive observables and differential studies in order to have better quantification for the formed medium properties, and accordingly better constraints for the medium parameters. Moreover, other interesting measurements that have shown similar phenomena in the high multiplicity events in proton-proton collisions as in central heavy ion collisions, e.g. long-range ridge-like structure \cite{13,14}, and strangeness enhancement \cite{15}, have stimulated the search for the onset of such and similar phenomena in the high multiplicity condition regardless of the colliding systems size. In addition, although it is well known that PYTHIA doesn't include final state interactions, previous simulated data (PYTHIA) \cite{13, 16} have shown qualitatively, but not quantitatively, similar patterns and structures for some of the observables, as in central heavy ion collisions. Such observations might indicate a non-trivial contributions for the commonly adopted observables from the underlying particle production mechanisms in QCD, and therefore the presented analysis.
In this analysis, PYTHIA has been used to simulate the p-p collisions, and the two-particle azimuthal correlations observable with its common metric quantity, the ratio between the yields as a function of colliding system size (i.e.; $I_{_{AA}}$ as an indicator for the parton energy loss in the medium), has been used in order to check for the onset of jet quenching in different multiplicity classes in p-p collisions.

In this article the quality assurance of the generated data, the two-particle angular correlations function, two-particle azimuthal correlations function, and the yield extractions are detailed and discussed in section 2; followed by the conclusions in section 3.

\section{Analysis and Results}

\subsection{Generated Data and Quality Assurance}
\label{startsample1}
\begin{figure}[!htbp]
\centering
      \resizebox{160mm}{200pt}{\includegraphics[width=160mm]{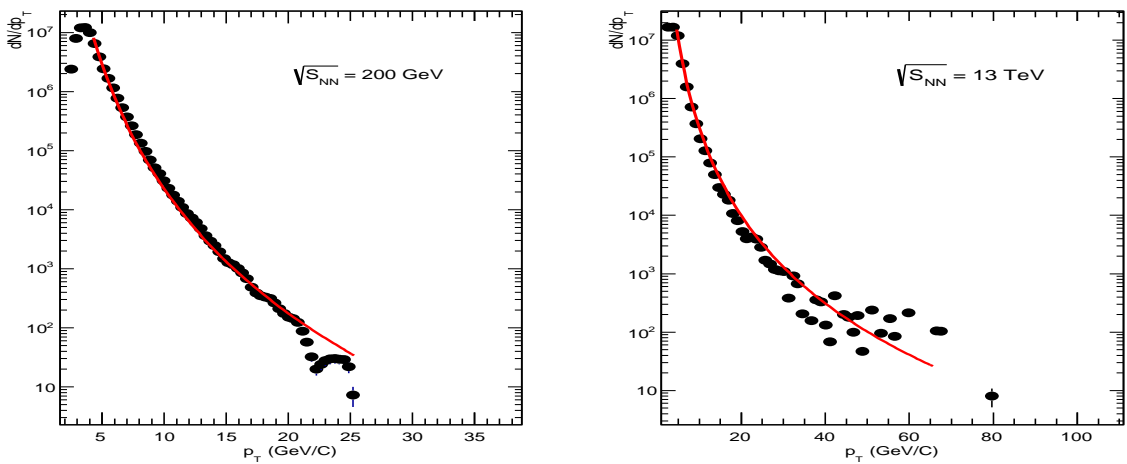}}
       \caption{Transverse momentum distributions of the produced particles with $p_{_T} > 3.0$ $GeV/c$, $|\eta|<2$, and $|\phi|<\pi$, at (left) $\sqrt{s_{_{NN}}} = 200$ $GeV$ and (right) $\sqrt{s_{_{NN}}} = 13 $ $TeV$. The red line show the power law fit with $1/p_{_{T}}^7$ at RHIC and with $1/p_{_{T}}^5$ at LHC energy.}
\label{fig:1}
\end{figure}

\begin{figure}[!htbp]
\centering
      \resizebox{160mm}{200pt}{\includegraphics[width=160mm]{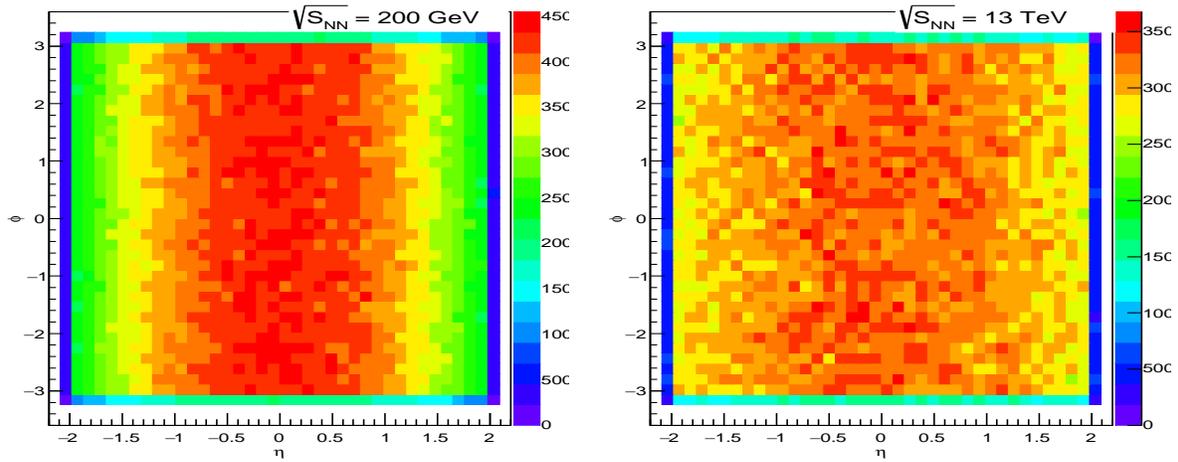}}
\caption{Pseudo rapidity-Azimuthal populations of the produced particles with $p_{_T} > 3.0$ $GeV/c$, $|\eta|<2$, and $|\phi|<\pi$, at (left) $\sqrt{s_{_{NN}}} = 200$ $GeV$ and (right) $\sqrt{s_{_{NN}}} = 13 $ $TeV$.}
\label{fig:2}
\end{figure}

The proton-proton collisions were simulated using PYTHIA 8 (version 8.185) \cite{17}, with its default parameters without being tuned to any of the studied collision energies, at two values of center of mass energies $\sqrt{s_{_{NN}}} = 200$ $GeV$ and $\sqrt{s_{_{NN}}} = 13$ $TeV$ corresponding to RHIC and LHC respectively. Around 30M events have been generated at each center of mass energy, where
the events selected for this analysis have at least one high transverse momentum particle, with $p_{_T} > 3.0$ $GeV/c$ (trigger) that is produced within the kinematic region of pseudo rapidity $|\eta|<2$ and full azimuth $|\phi|<\pi$. The kinematic cuts have been imposed in this analysis are to resemble the detectors coverages and capabilities at the experiments. Figure \ref{fig:1} shows the transverse momentum distributions of the trigger particles, and as expected the particle distribution is rapidly falling at RHIC than at LHC, and accordingly the distributions exhibit more kinematics reach at higher center of mass energy. The ($dN/dp_{_{T}}$) distributions on $p_{_{T}}$ have shown reasonable power-law fit, within feasible range, with $1/p_{_{T}}^7$ at RHIC and with $1/p_{_{T}}^5$ at LHC energy. Figure \ref{fig:2} represents the populations of the trigger particles in the pseudo rapidity and azimuthal directions. As shown the pattern of the distribution is approximately uniform within the covered kinematic regions. The uniform distributions of the trigger particles in the azimuth are the most relevant ones to this analysis, which assure that there is no bias in the azimuthal direction.

\subsection{Two-particle Angular Correlations function}
\label{startsample2}
With respect to each trigger particle, with $p_{_T} > 3 GeV/c$, $|\eta|< 2$ and $|\phi|<\pi$, the relative azimuthal and pseudo rapidity positions of all of the other produced particles (associated), with transverse momentum less than the trigger particle’s transverse momentum ($p_{_{T}}^{assoc} < p_{_{T}}^{trg}$) in the same event within $|\eta|< 2$, and $|\phi|<\pi$, are measured. The difference in azimuthal angle and pseudo rapidity are shown in Fig. 3 as $\Delta \phi$ vs. $\Delta \eta$, where $\Delta \phi$ = $\phi^{trg}$ - $\phi^{assoc}$, and $\Delta\eta$ = $\eta^{trg}$ - $\eta^{assoc}$.
\begin{figure}[!htbp]
\centering
       \resizebox{160mm}{200pt}{\includegraphics[width=160mm]{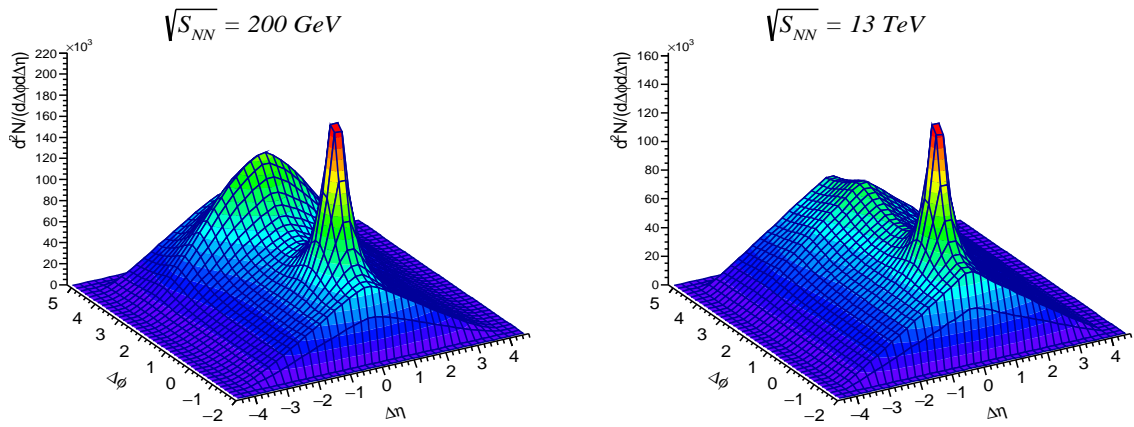}}
      \caption{Two-particle correlation functions $d^{2}N/d(\Delta\eta\Delta\phi)$ versus $\Delta \eta$, and $\Delta \phi$ at (left) $\sqrt{s_{_{NN}}} = 200$ $GeV$ and (right) $\sqrt{s_{_{NN}}} = 13 $ $TeV$.}
\label{fig:3}
\end{figure}
The two-dimensional correlations structure shown in Fig. 3 exhibits: 1) a narrow peak at ($\Delta \eta$, $\Delta \phi$) $\approx$ (0, 0) which can be understood as the contribution from higher $p_{_{T}}$ clusters (e.g., hard processes like jets). As expected from the basic principles of QCD and its confinement features, that few of the produced particles are originated from the same parent parton, and are accordingly correlated (i.e., near-side yield); 2) a ridge at $\Delta \phi$ $\approx$ $\pi$ that spans a wide range in $\Delta \eta$, which can be regarded as away-side jets due to the momentum conservation; and 3) an approximately Gaussian ridge at $\Delta \eta$ $\approx$ 0 which becomes broader as
\begin{figure}[!htbp]
\centering
     \resizebox{160mm}{200pt}{\includegraphics[width=160mm]{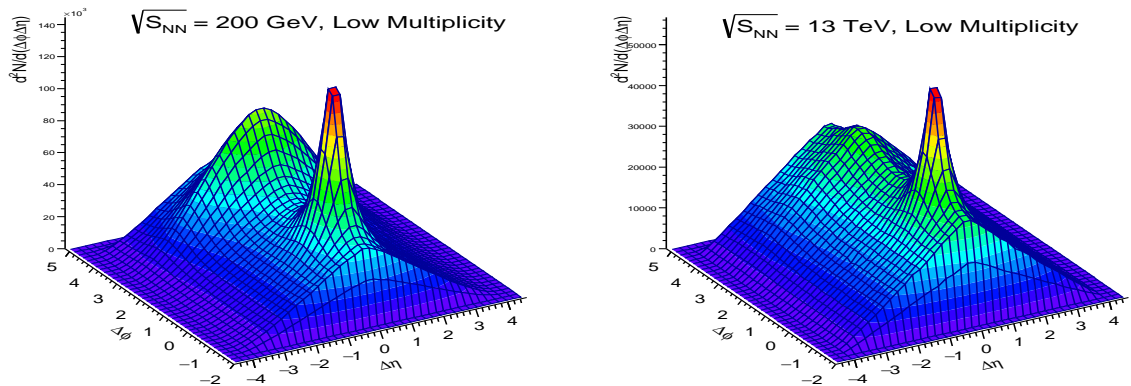}}
      \caption{Two-particle correlation functions $d^{2}N/d(\Delta\eta\Delta\phi)$ versus $\Delta \eta$, and $\Delta \phi$ for the low multiplicity events (0 $<$ mult $\leq$ 20) at (left) $\sqrt{s_{_{NN}}} = 200$ $GeV$ and (right) $\sqrt{s_{_{NN}}} = 13 $ $TeV$.}
\label{fig:4}
\end{figure}
\begin{figure}[!htbp]
\centering
      \resizebox{160mm}{200pt}{\includegraphics[width=160mm]{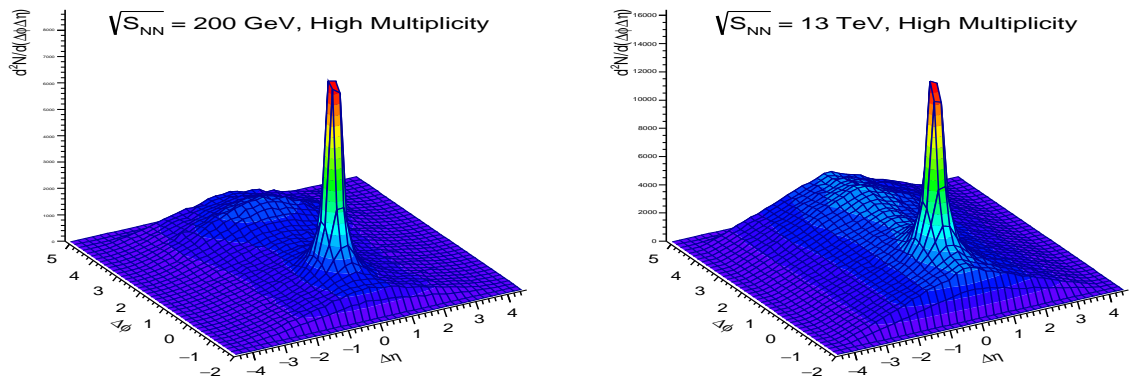}}
      \caption{Two-particle correlation functions $d^{2}N/d(\Delta\eta\Delta\phi)$ versus $\Delta \eta$, and $\Delta \phi$ for the high multiplicity events (RHIC: mult $\geq$ 40, LHC: 0 mult $\geq$ 120) at (left) $\sqrt{s_{_{NN}}} = 200$ $GeV$ and (right) $\sqrt{s_{_{NN}}} = 13 $ $TeV$.}
\label{fig:5}
\end{figure}
$\Delta \phi$ increases, resulting from the decay of lower $p_{T}$ clusters. The obvious difference of the level of backgrounds, and the relative strength between the near-side and away-side yields at
$\sqrt{s_{_{NN}}} = 200$ $GeV$ and $\sqrt{s_{_{NN}}} = 13 $ $TeV$ are mainly due to the different center of mass energy and accordingly different probed regions for the parton distribution functions, and accordingly different fragmentation, at the same kinematics.
Figures 4 and 5, show the evolution of the two-dimensional correlations function on the multiplicity at $\sqrt{s_{_{NN}}} = 200$ $GeV$ and $\sqrt{s_{_{NN}}} = 13 $ $TeV$. As it is clearly shown, the relative strength of the away-side to the near-side is dramatically suppressed in the high multiplicity events (low multiplicity at both energies: 0 $<$ mult. $\leq$ 20; high multiplicity: RHIC mult. $\geq$ 40, LHC mult. $\geq$ 120), at both energies. As it was suggested by Bjorken \cite{18} that parton traversing bulk partonic matter undergoes significant energy loss, with observable consequences on the parton's subsequent fragmentation into hadrons, indeed the previously reported results \cite{19, 20} have shown the away-side to be increasingly suppressed with centrality. In order to quantify such observations, the near and away-side yields have been calculated per each trigger particle for each $z_{_{T}}$ bin ($z_{_{T}} = p_{_T}^{assoc} / p_{_T}^{trig}$), for each multiplicity event class at both energies, as detailed at the next section.
 \begin{figure}[!htbp]
\centering
   \resizebox{160mm}{200pt}{\includegraphics[width=160mm]{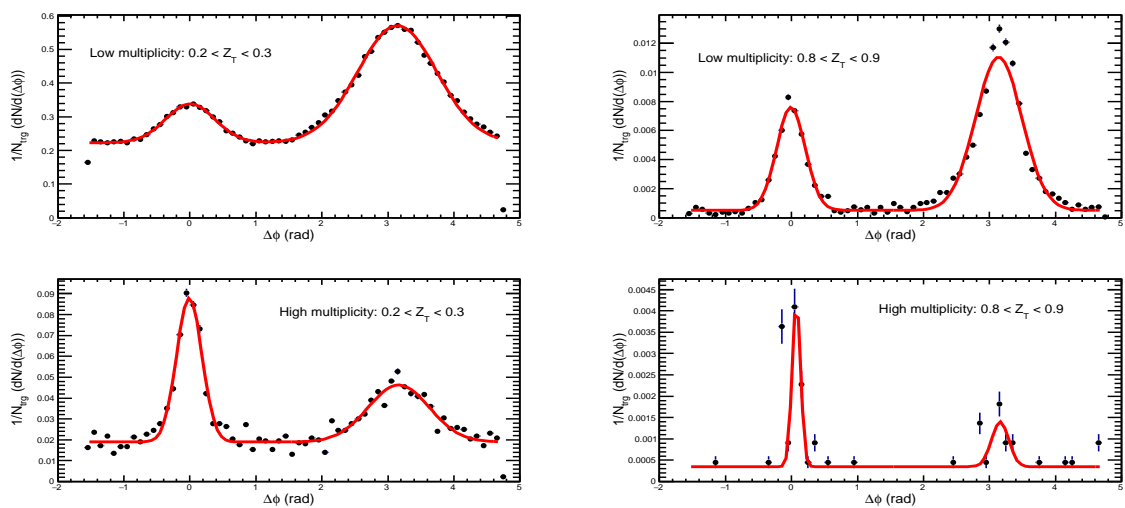}}
 \caption{Two-particle azimuthal correlation functions normalized per trigger particle $1/N_{trg} (dN/d(\Delta\phi))$ versus $\Delta \phi$ for similar $z_{_{T}}$ values of the low (upper) and high (lower) multiplicity events at $\sqrt{s_{_{NN}}} = 200$ $GeV$. The fit consist of two-Gaussian peaks and straight line in order to estimate the level of backgrounds.}
\label{fig:6}
\end{figure}
\begin{figure}[!htbp]
\centering
      \resizebox{160mm}{200pt}{\includegraphics[width=160mm]{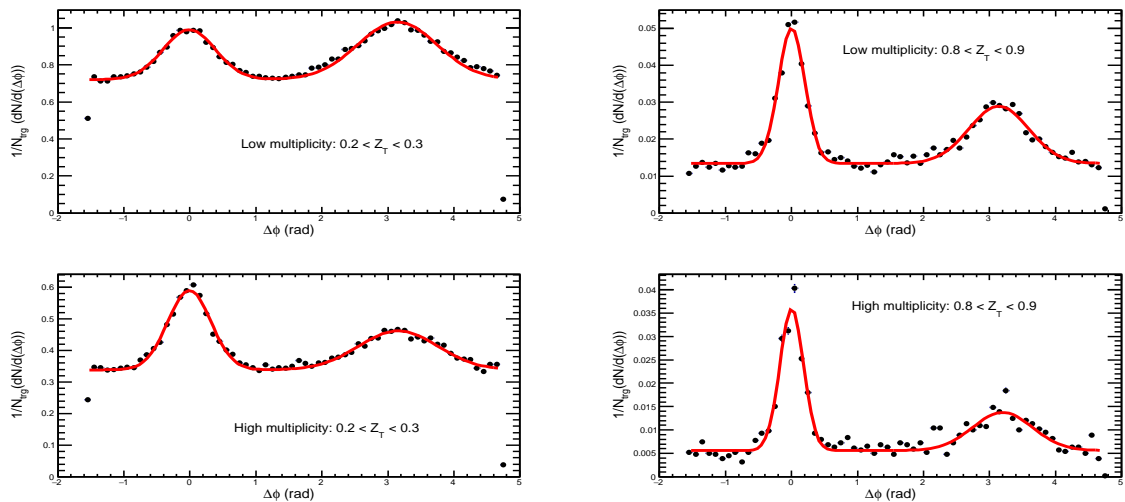}}
\caption{Two-particle azimuthal correlation functions normalized per trigger particle $1/N_{trg} (dN/d(\Delta\phi))$ versus $\Delta \phi$ for similar $z_{_{T}}$ values of the low (upper) and high (lower) multiplicity events at $\sqrt{s_{_{NN}}} = 13 $ $TeV$. The fit consist of two-Gaussian peaks and straight line in order to estimate the level of backgrounds.}
\label{fig:7}
\end{figure}
\subsection{Two-particle Azimuthal Correlations}
\label{startsample3}
In order to search for the onset of the jet quenching, the two dimensional correlation functions were reduced to one-dimensional functions of $\Delta\phi$ integrated over the entire range of $|\Delta\eta|$ $\le$ 2. The two-particle azimuthal correlation functions are normalized by the number of trigger particles on event-by-event basis for the low and high multiplicity events at both energies. Figures 6, and 7 show sample of the two-particle azimuthal correlation functions for similar $z_{_T}$ values of the low and high multiplicity at both energies. The $z_{_T}$ measures the relative energy in terms of the trigger particle, since this analysis does not reconstruct the full jet. One of the common features of all azimuthal correlation functions for each $z_{_T}$ bin, for low and high multiplicity events at the different center-of-mass energies, is that the levels of uncorrelated background particles are strongly suppressed with increasing the $z_{_T}$ values which indicates that the high-$p_{_{T}}$ particles have resulted from the hard-scattering (jet-like events).
It is obvious that the relative strength of the away-side to the near side, and the peak widths change with multiplicity as well as with $z_{_T}$. In order to quantify the multiplicity effects on the near and away sides, the yields at both sides have been extracted for each $z_{_{T}}$ bin of 0.1 width at both energies as explained in the next section.
\begin{figure}[!htbp]
\centering
      \subfloat[]{\includegraphics[width=80mm]{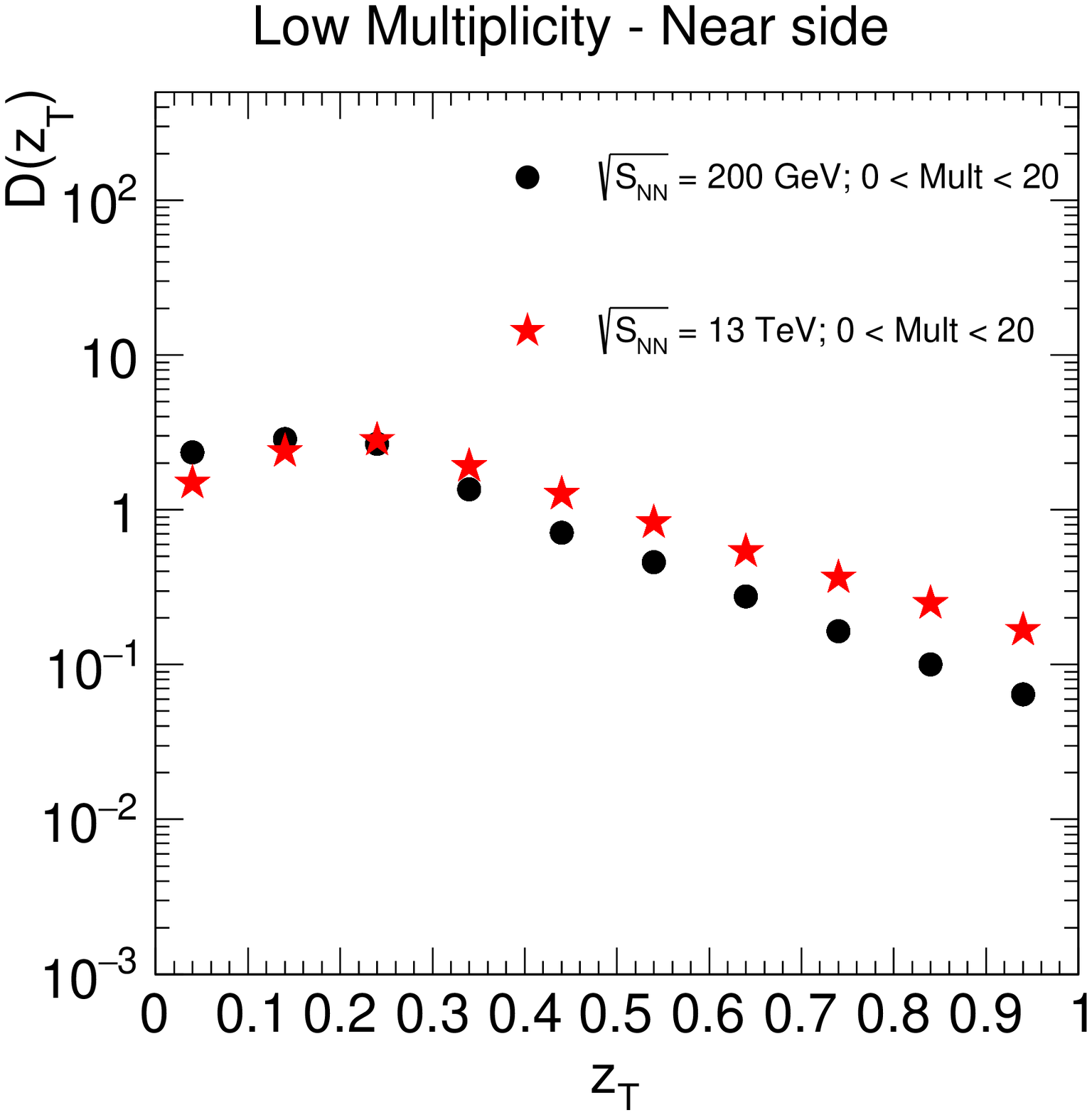}}
      \subfloat[]{\includegraphics[width=80mm]{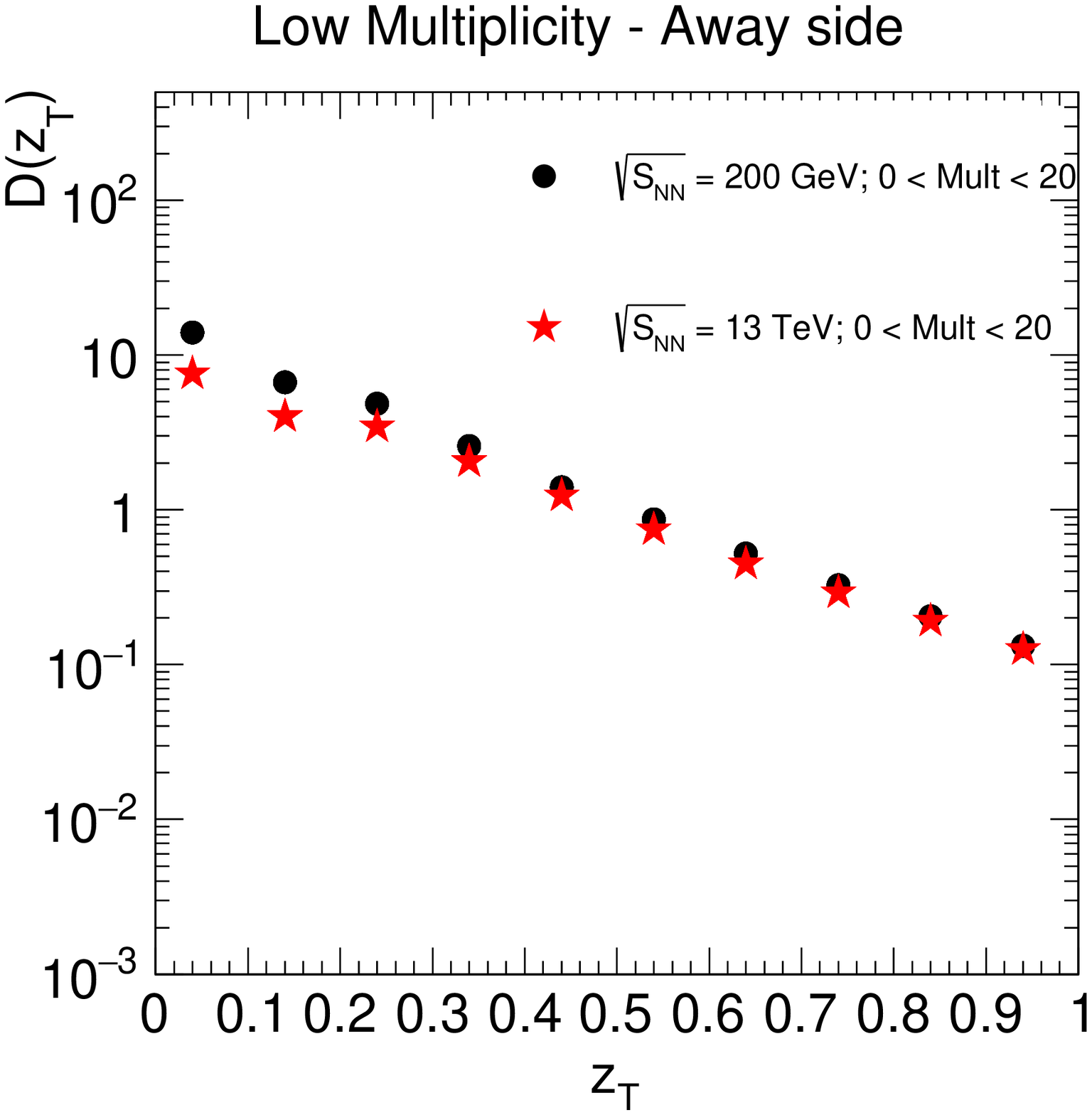}}
\caption{The $z_{T}$ dependence of the yield per trigger for the near-side (a), and away-side (b) of the low multiplicity events at $\sqrt{s_{_{NN}}} = 200$ $GeV$ and $\sqrt{s_{_{NN}}} = 13 $ $TeV$.}
\label{fig:8}
\end{figure}
\subsection{Yield Extractions}	
The number of associated particles per trigger, i.e., the yield per trigger, $D(z_{_{T}})$, where $D(z_{_{T}}) = \frac{1}{N_{trg}}\frac{dN}{d(\Delta\phi)}$, is found by counting the entries underneath the peaks, without the fit, within certain $|\Delta\phi|$ window after subtracting the background as determined from the straight line of the fit at the near side and the away side for each $z_{T}$ bin, as shown in Fig. 6 and 7. The integrating region for near side yield was $|\Delta\phi| < 0.63$ and for the away side was $|\Delta\phi-\pi| < 0.63$.
\begin{figure}[!htbp]
\centering
       \subfloat[]{\includegraphics[width=80mm]{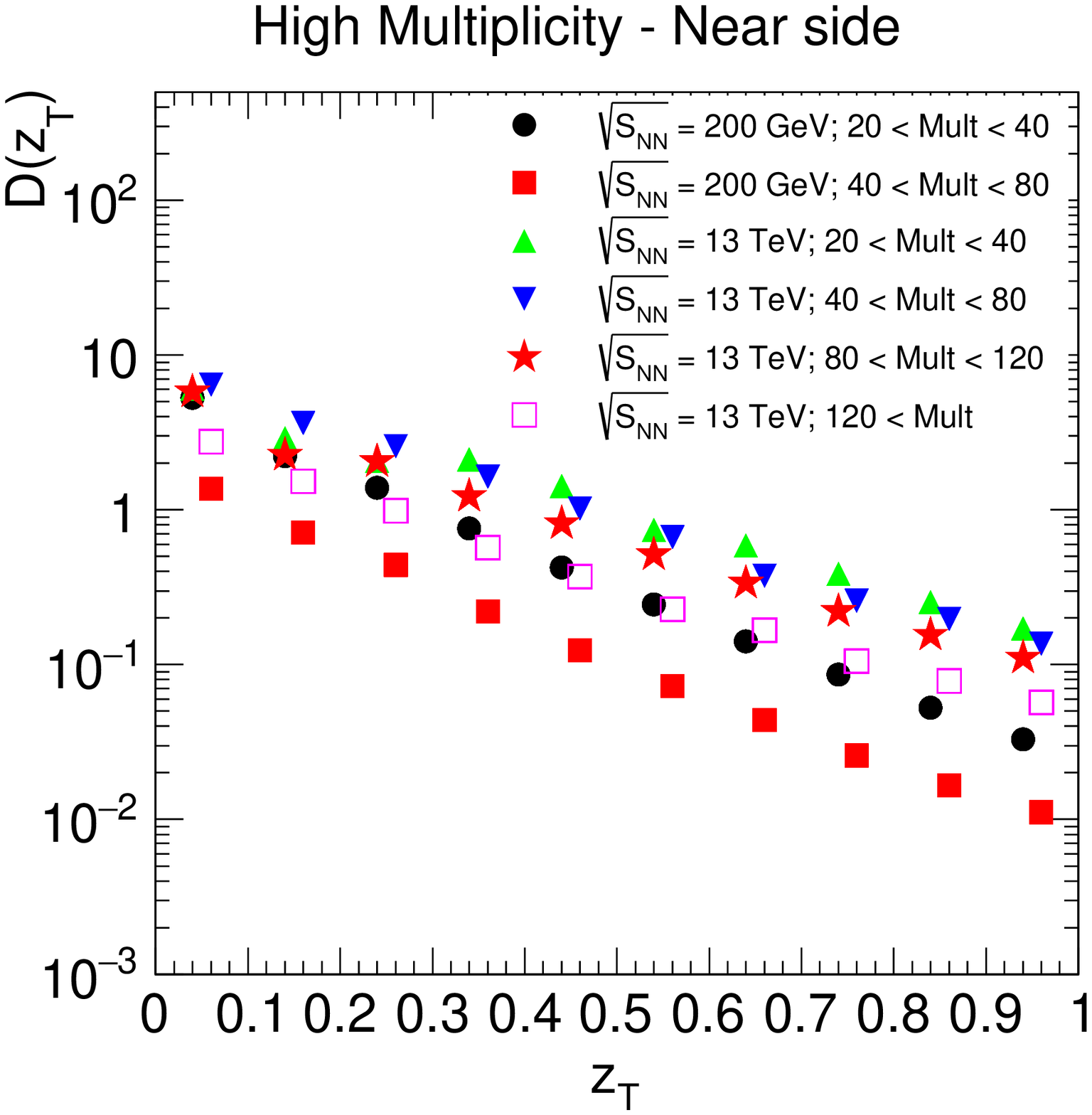}}
       \subfloat[]{\includegraphics[width=80mm]{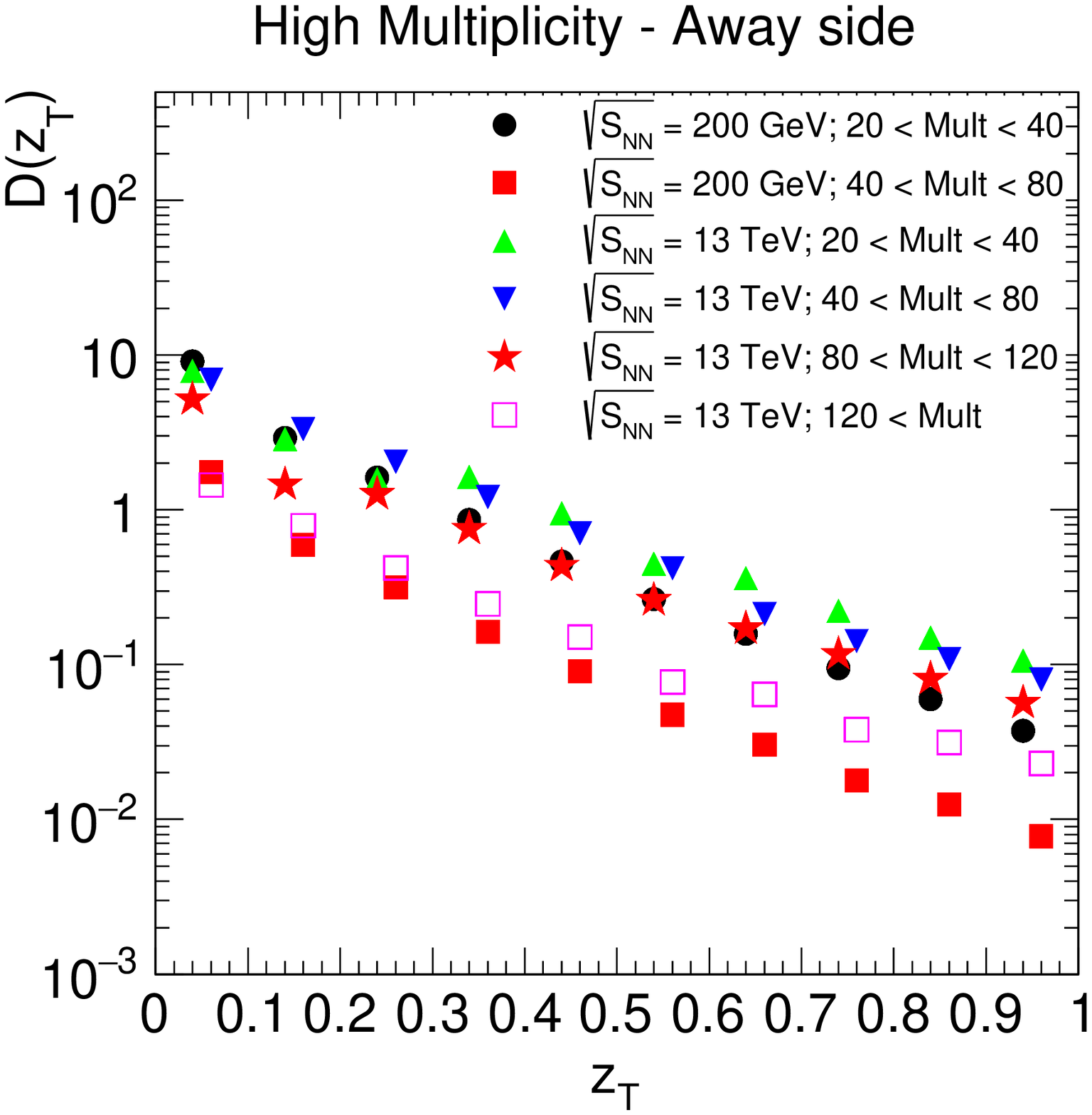}}
\caption{The $z_{T}$ dependence of the yield per trigger for the near-side (a), and away-side (b) of the high multiplicity events at $\sqrt{s_{_{NN}}} = 200$ $GeV$ and $\sqrt{s_{_{NN}}} = 13 $ $TeV$.}
\label{fig:9}
\end{figure}
For each $z_T$ bin, the normalized near side and away side yields were extracted at the RHIC and LHC corresponding energies for the low and high multiplicity classes, as shown in figures 8 and 9. RHIC data has been classified into three bins of multiplicity: 0 - 20, 20 - 40, and 40 - 80, while LHC data provides enough statistics for five multiplicity bins: 0 - 20, 20 - 40, 40 - 80, 80 - 120, and greater than 120 particles. The near side yields at both energies for the low multiplicity events show an unexpected pattern at low $z_{_{T}}$, as a reflection for the cut-off momentum for the string fragmentation in PYTHIA calculations. As it is clearly shown at Figures 8 and 9, all yields at near- and away-sides are dropping as a function of $z_{_{T}}$, which can be understood in the light of the steeply falling parton distribution functions. At both energies the way-side yield is greater than the near-side for the low multiplicity events, Fig. 8, which can be attributed to the trigger bias. However such difference between the near- and away-sides is not significant, if any, for the high multiplicity events at both energies. Also, the near-side and away-side yields are very similar at different energies for the low multiplicity, but they show systematic differences for the high multiplicity events.
In order to quantify the multiplicity effects, if any, the ratio between the near-side yields at high and low multiplicities ($I_{_{HL}}^{^{N}}$),
and away-side yields at high and low multiplicities ($I_{_{HL}}^{^{A}}$), where:
\begin{equation}
\it{I}_{_{HL}}^{^{N}} (z_{_{T}}) = \frac {D^{^{high-mult}}_{_{near-side}}(z_{_{T}})} {^{D^{low-mult}}_{{near-side}}(z_{_{T}})}; \hspace{2.0 cm}
I_{_{HL}}^{^{A}} (z_{_{T}}) = \frac {D^{^{high-mult}}_{_{away-side}}(z_{_{T}})} {^{D^{low-mult}}_{{away-side}}(z_{_{T}})}
\end{equation}

have been calculated and plotted for both energies, as a function of $z_{T}$, as shown in Figure 10.
\begin{figure}[!htbp]
\centering
       \subfloat[]{\includegraphics[width=80mm]{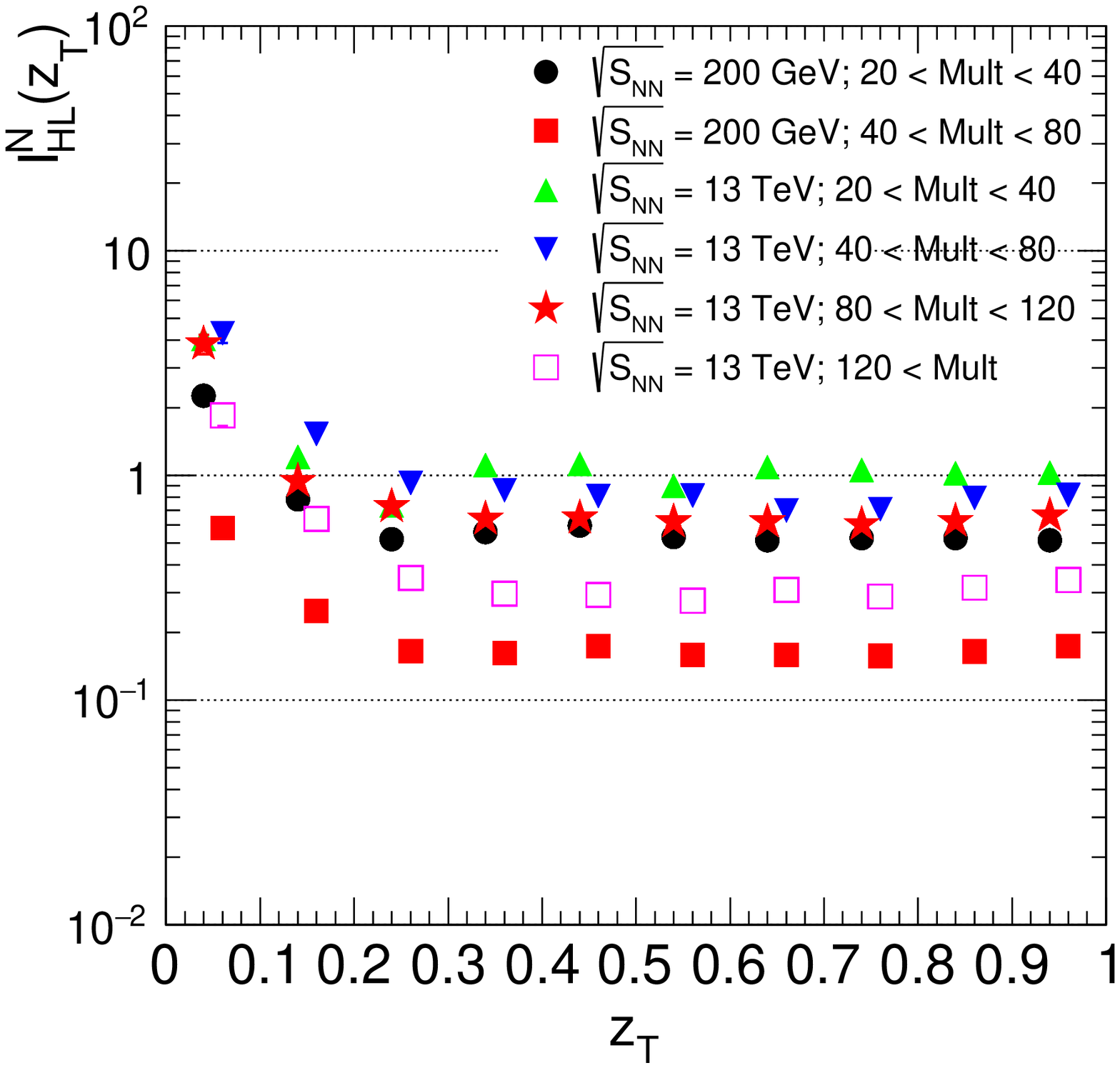}}
       \subfloat[]{\includegraphics[width=80mm]{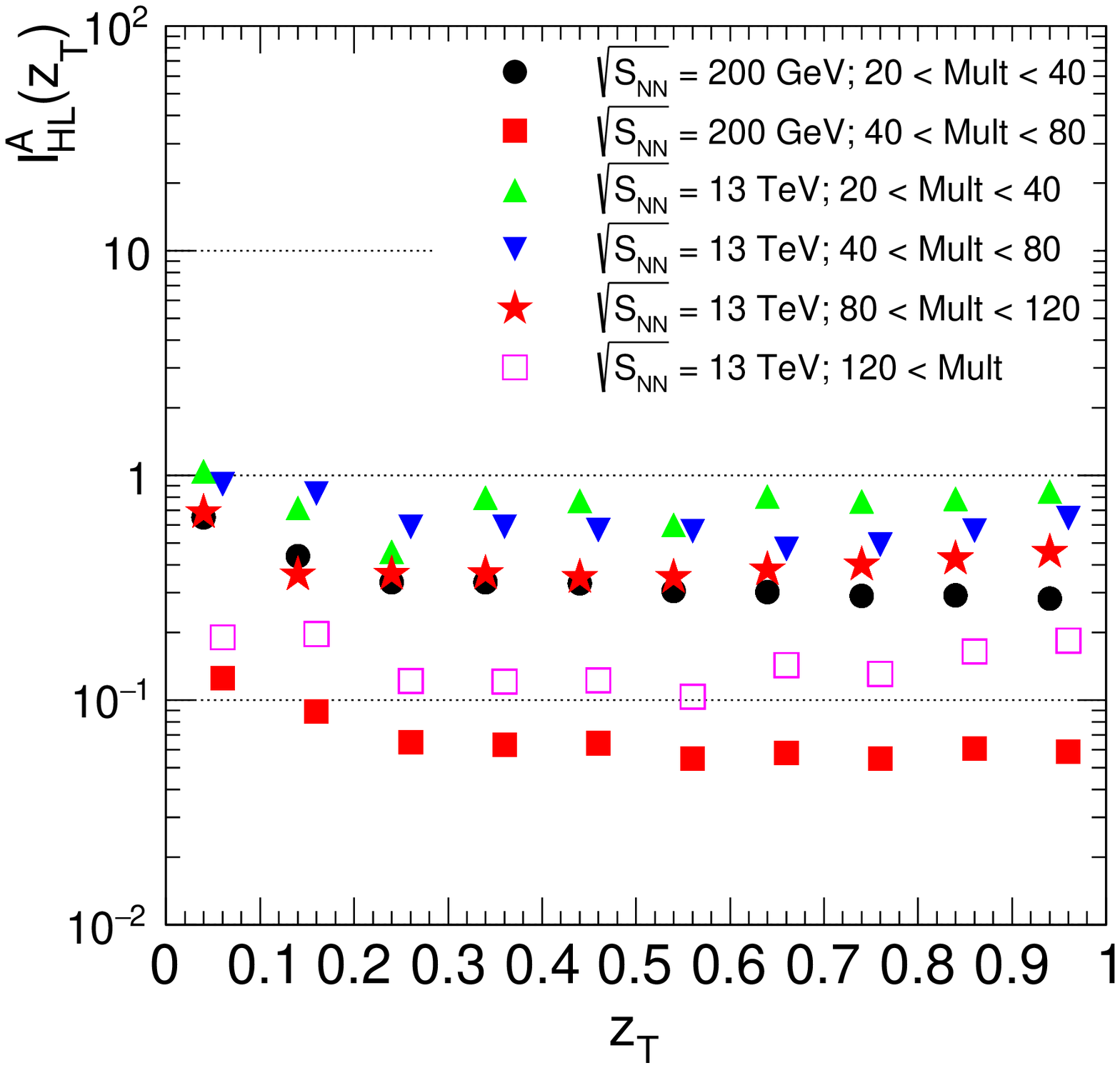}}
\caption{The $z_{_{T}}$ dependence of (a) $I_{_{HL}}^{^{N}}$ and (b) $I_{_{HL}}^{^{A}}$ for different multiplicity at $\sqrt{s_{_{NN}}} = 200$ $GeV$ and $\sqrt{s_{_{NN}}} = 13 $ $TeV$.}
\label{fig:10}
\end{figure}
The suppression of both yields show no strong dependence on $z_{_{T}}$, which agrees with the previously reported results \cite{12}, and accordingly the suppression is independent of the initial parton energy within the covered kinematics range. As it is shown at figure 10, indeed there is a multiplicity dependence for the ratios at both sides, near and away. The values of $I_{_{HL}}^{^{N}}$, and $I_{_{HL}}^{^{A}}$ are always less than unity except for the near side at low $z_{_{T}}$, which is due to an artificial effect of the lower momentum cuts of PYTHIA for the string fragmentation. Both quantities are continuously decreasing with increasing the multiplicity at each center of mass energy, with more suppression at RHIC energy for the same multiplicity, which might be due to different probed region of the parton distribution functions. It is also noticeable that the multiplicity effect on the away side ratios is more than its effect on the near side ratios, which might be due to 1) the combinations of the trigger bias at the near side, 2) different types of partons at the away-side than at the near sides, as it is known that the quark fragmentation is harder than those of gluons.
\section{Conclusions}
The multiplicity has an obvious effect on the ratios of near-side and away-side yields, with more suppression at the away-side than these at the near-side. The suppression in the yields does not exhibit a parton energy dependence within the covered kinematics range. The near and away-side yields show stronger dependence on the center-of-mass energy for the high multiplicity events but not for the low multiplicity events. According to the presented analysis, the multiplicity dependence of the yield suppressions cannot be used neither to exclude the jet quenching nor to prove it in the high multiplicity events in p-p collisions, as the suppressions have been found at both sides, near and away of the high-$p_{_{T}}$ particles, within the reported kinematics region. The fact that the near-side shows a suppression in the high multiplicity events in p-p collisions is not consistent with the surface bias emission as reported by various experiments, and accordingly such suppression at both sides shown in this analysis could be due to 1) the evolution of the parton distribution functions on the trigger particle momentum, 2) parton energy loss in the high multiplicity events, or 3) a combination of both effects 1) and 2).
More future studies at higher transverse momenta could be used to either rule out or to confirm the multiplicities dependence for the jet quenching commonly adopted observables.
\ack
The authors would like to thank the American University in Cairo for supporting this research.


\section*{References}
\begin{thebibliography}{9}
\bibitem{1} Adams J \etal [STAR collaboration] 2005 {\it Null. Phys. A} {\bf 757} 232301 
\bibitem{2} Adcox K \etal [PHOENIX collaboration] 2005 {\it Null. Phys. A} {\bf 757}
\bibitem{3} Arlene I \etal [BRAHMS collaboration] 2005 {\it Null. Phys. A} {\bf 757}
\bibitem{4} Back B B \etal [PHOBOS collaboration] 2005 {\it Null. Phys. A} {\bf 757}
\bibitem{5} Chatrchyan S \etal [CMS collaboration] 2012 {\it Eur. Phys. J. C} {\bf 72} 1945 
\bibitem{6} Adler S S \etal [PHOENIX collaboration] 2005 \PRL {\bf 94} 232301 
\bibitem{7} Chatrchyan S \etal [CMS collaboration] 2012 {\it Phys. Lett. B} {\bf 710} 256 
\bibitem{8} Chatrchyan S \etal [CMS Collaboration] 2012 {\it Phys. Lett. B} {\bf 715} 66 
\bibitem{9} Chatrchyan S \etal [CMS collaboration] 2011 \PRL {\bf 106} 212301 
\bibitem{10} Chatrchyan S \etal [CMS collaboration] 2012 \JHEP JHEP05(2012)063
\bibitem{11} Abelev B I \etal [ALICE Collaboration] 2012 \JHEP JHEP09(2012)112 
\bibitem{12} Abelev B I \etal [STAR collaboration] 2010 {\it Phys. Rev. C} 34909 
\bibitem{13} Alver B \etal [PHOBOS collaboration] 2007 {\it Phys. Rev. C} {\bf 75} 054913
\bibitem{14} Khachatryan V \etal [CMS Collaboration] 2010 \JHEP JHEP09(2010)091 
\bibitem{15} Adam J \etal [ALICE Collaboration] 2017 {\it Nature Physics} {\bf 13} 535
\bibitem{16} “Multiple Parton Interactions at the LHC. Proceedings, 1st Workshop, Perugia, Italy, October 27-31, 2008”. DESY-PROC-2009-06
\bibitem{17} http://home.thep.lu.se/$\sim$torbjorn/Pythia.html
\bibitem{18} Bjorken J D, FERMILAB-PUB-82-59- THY and erratum (unpublished)
\bibitem{19} Adams J \etal [STAR collaboration] 2003 {\it Phys. Rev. Lett.} {\bf 91} 072304
\bibitem{20} Adams J \etal [STAR collaboration] 2006 {\it Phys. Rev. Lett.} {\bf 97} 162301

\end{thebibliography}
\end{document}